\documentclass[amsmath,amssymb,showpacs,twocolumn]{revtex4}
\usepackage{graphicx}
\makeatletter \@addtoreset{equation}{section} \makeatother
\renewcommand{\theequation}{\arabic{section}.\arabic{equation}}
\newcommand{\disp}{\displaystyle}
\def\ri{{\rm i}}
\def\rd{{\rm d}}
\def\re{{\rm e}}
\def\rn{{\rm n}}

\newcommand{\vectorTwo}[2]{\left( \begin{array}{c} #1 \\ #2 \end{array} \right) }

\begin{document}
\title{Stability analysis for solitons in PT-symmetric optical lattices}
\author{Sean Nixon$^{1}$, Lijuan Ge$^{1,2}$, and
Jianke Yang$^{1}$ \hspace{-0.35cm} \footnote{Corresponding author,
email: jyang@math.uvm.edu}}
\affiliation{%
$^{1}$ Department of Mathematics and Statistics, University of
Vermont, Burlington, VT 05401, USA \\
$^{2}$ Department of Physics, Shanghai University, China
}%

\begin{abstract}

Stability of solitons in parity-time (PT)-symmetric periodic
potentials (optical lattices) is analyzed in both one- and
two-dimensional systems. First we show analytically that when the
strength of the gain-loss component in the PT lattice rises above a
certain threshold (phase-transition point), an infinite number of
linear Bloch bands turn complex simultaneously. Second, we show that
while stable families of solitons can exist in PT lattices,
increasing the gain-loss component has an overall destabilizing
effect on soliton propagation. Specifically, when the gain-loss
component increases, the parameter range of stable solitons shrinks
as new regions of instability appear. Thirdly, we investigate the
nonlinear evolution of unstable PT solitons under perturbations, and
show that the energy of perturbed solitons can grow unbounded even
though the PT lattice is below the phase transition point.

\end{abstract}


\maketitle

\section{Introduction}

Recent interest in study of parity-time (PT) symmetric optical
potentials has its roots in quantum mechanics. In quantum mechanics,
in order for the energy levels to be real and the theory to be
probability conserving, it is usually assumed that the Hamiltonian
(Schr\"odinger) operator be Hermitian. However, in the past decade
there has seen considerable attention \cite{Bender1998, Bender2002,
Ahmed2001, Mostafazadeh2003} in a weaker version of the Hermiticity
axiom which requires that the Hamiltonian instead only exhibit
space-time reflection symmetry (PT symmetry). While there has been
much theoretical success in developing a non-Hermitian quantum field
theory, the phenomena unique to this class of pseudo-Hermitian
systems have not yet been observed experimentally.

The same Schr\"{o}dinger equation from quantum mechanics applies
also to optics. Motivated by this connection, optical systems which
have PT-symmetric potentials have been formulated
\cite{Christodoulides2007}. A PT-symmetric optical potential
$V(\textbf{x})$ is realizable by the careful distribution of gain
and loss in the media so that it satisfies the PT symmetry
$V(\textbf{x}) = V^*(-\textbf{x})$, where $\textbf{x}$ is the
spatial coordinate and `$*$' stands for complex conjugation. That
is, the refractive-index profile of the media is even and gain-loss
profile is odd. Such optical PT media have been created
experimentally \cite{Guo2009, Segev2010}. These linear PT media
undergo phase transition as the gain-loss component crosses a
certain threshold \cite{Bender1998,Musslimani2008,Guo2009,
Segev2010}. Below this threshold, all eigenvalues of the PT
potential are real; but above this threshold, complex eigenvalues
appear, hence the intensity of a light beam grows exponentially
during linear propagation. The nature of this phase transition
(especially for periodic PT potentials) has not been fully
understood yet.

These phenomena may also be studied in a nonlinear context by
considering the existence of localized modes called solitons
\cite{Musslimani2008,Musslimani2011}. When a system contains gain
and loss, solitons generally exist only at special values of the
propagation constant \cite{Akhmediev_book}. However, since PT
potentials can admit all-real linear spectra, solitons could exist
at continuous ranges of the propagation constant, which is quite
remarkable. So far, soliton families in PT-symmetric periodic
potentials with defects and in PT-symmetric nonlinear potentials
have been investigated
\cite{Wang2011,Zhang2011,Konotop2011,KonotopArxiv}. But stability
properties of these PT solitons (especially in periodic PT
potentials) have not been carefully examined.

In this paper, we investigate linear phase transition and stability
of (nonlinear) solitons in PT-symmetric periodic potentials (optical
lattices) in both one and two spatial dimensions. Our mathematical
model is the nonlinear Schr\"{o}dinger (NLS) equation with a PT
lattice potential,
\begin{equation} \label{e:model}
i U_z + U_{xx} + U_{yy} + V(x,y)U + \sigma |U|^2U = 0,
\end{equation}
where $\sigma = \pm 1$ denotes the focusing and defocusing
nonlinearity, and the potential $V(x,y)$ is periodic in $x$ and $y$
and satisfies the PT symmetry $V(x,y) = V^*(-x,-y)$. For simplicity,
we take this PT lattice potential to be
\begin{equation}  \label{e:PTlattice}
V(x) = V_0\left[ \cos^2(x) + \ri W_0 \sin(2x) \right]
\end{equation}
in one dimension (1D) and
\begin{equation} \label{e:PTlattice2D}
V(x,y) = V_0 \left\{\cos^2(x) + \cos^2(y) + \ri W_0\left[\sin(2x) +
\sin(2y) \right]\right\}
\end{equation}
in two dimensions. Here $V_0 \hspace{0.05cm} (\hspace{0.04cm}>0
\hspace{0.03cm})$ is the depth of the real component of the
potential, $W_0$ is the relative magnitude of the imaginary
component, and the period of this PT lattice is $\pi$. For this
system, we first show analytically that when the strength of the
gain-loss component (the imaginary part of $V(x,y)$) in the PT
lattice rises above a certain threshold (phase-transition point), an
infinite number of linear Bloch bands turn complex simultaneously.
This simultaneous bifurcation of an infinite number of complex
eigenvalues at the phase transition point has never been reported
before for any PT-symmetric potentials to our best knowledge
\cite{Bender1998,Musslimani2011}. Second, we show that while stable
families of solitons can exist in PT lattices (below the phase
transition point), increasing the gain-loss component has an overall
destabilizing effect on soliton propagation. Specifically, when the
gain-loss component increases, the parameter range of stable
solitons shrinks as new regions of instability appear. Thirdly, we
investigate the evolution of unstable PT solitons under
perturbations, and show that the energy of these perturbed solitons
can grow unbounded even if the PT lattice is below the phase
transition point.

\section{Simultaneous complex-eigenvalue bifurcation at the phase transition point}
\label{Sec:PhaseTransition}

We begin by investigating the bifurcation of the continuous spectrum
(Bloch bands) of the linear Schr\"odinger operator in Eq.
(\ref{e:model}) at the phase transition point. The phase transition
point is a point above which the spectrum is no longer purely real
as the strength of the imaginary (gain-loss) contribution relative
to the real (refractive-index) contribution in the potential is
increased. We will show that at the phase transition point, an
infinite number of Bloch bands turn complex simultaneously.

We first consider this bifurcation in one dimension. In this case,
the linear Schr\"odinger equation
is
\begin{equation} \label{e:linear1D}
i U_z + U_{xx} + V(x)U = 0,
\end{equation}
where the PT lattice potential $V(x)$ is given in Eq.
(\ref{e:PTlattice}).
The continuous spectrum of this Schr\"odinger equation consists of
Bloch modes of the form
\begin{equation} \label{e:Bloch1D}
U(x,z)=p(x; k)e^{ikx-i\mu z},
\end{equation}
where $p(x;k)$ is a $\pi$-periodic function in $x$, $k$ is the
wavenumber in the irreducible Brillouin zone $-1\le k\le 1$, and
$\mu$ is the propagation constant. The values of $\mu$ and $k$ are
related. The relation $\mu=\mu(k)$ is called the diffraction
relation, and all admissible values of $\mu$ form the continuous
spectrum of Eq. (\ref{e:linear1D}).

\begin{figure}[!htbp]
\centering
\includegraphics[width=0.45\textwidth]{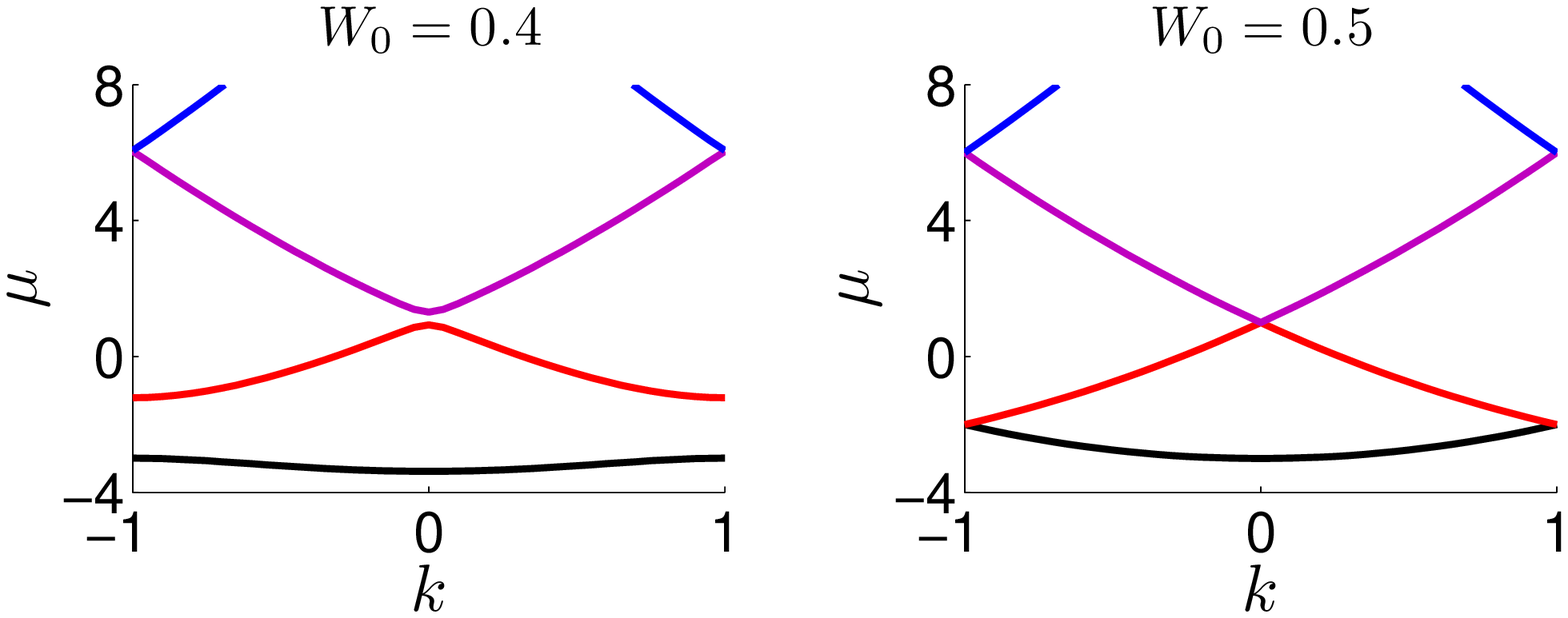}

\vspace{0.5cm}
\includegraphics[width=0.45\textwidth]{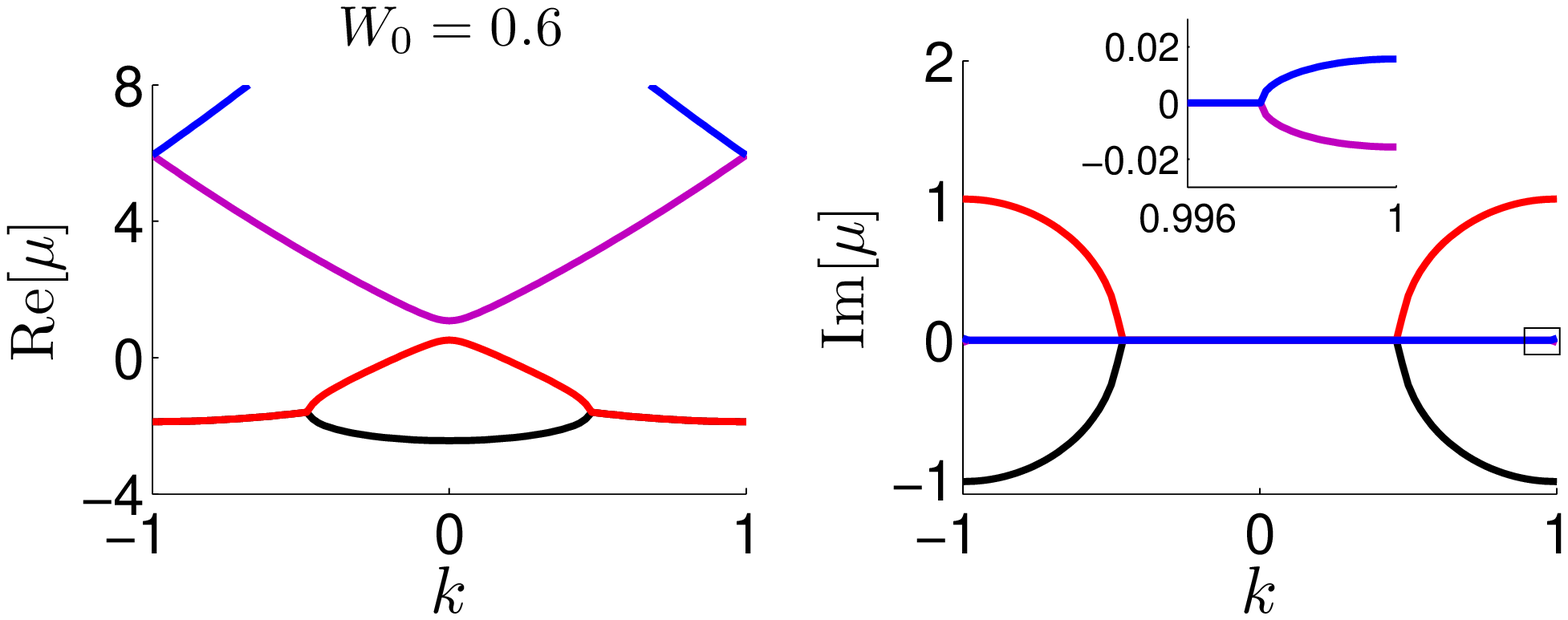}
\caption{(Color online) Diffraction relations of PT lattices
(\ref{e:PTlattice}) for three $W_0$ values 0.4, 0.5 (upper panel)
and 0.6 (lower panel) at $V_0=6$. The inset in the lower right panel
is amplification of the small boxed region near $k=1$ and
$\mbox{Im}[\mu]=0$ of the same panel. } \label{f:fig1}
\end{figure}

\begin{figure}[!htbp]
\centering
\includegraphics[width=0.4\textwidth]{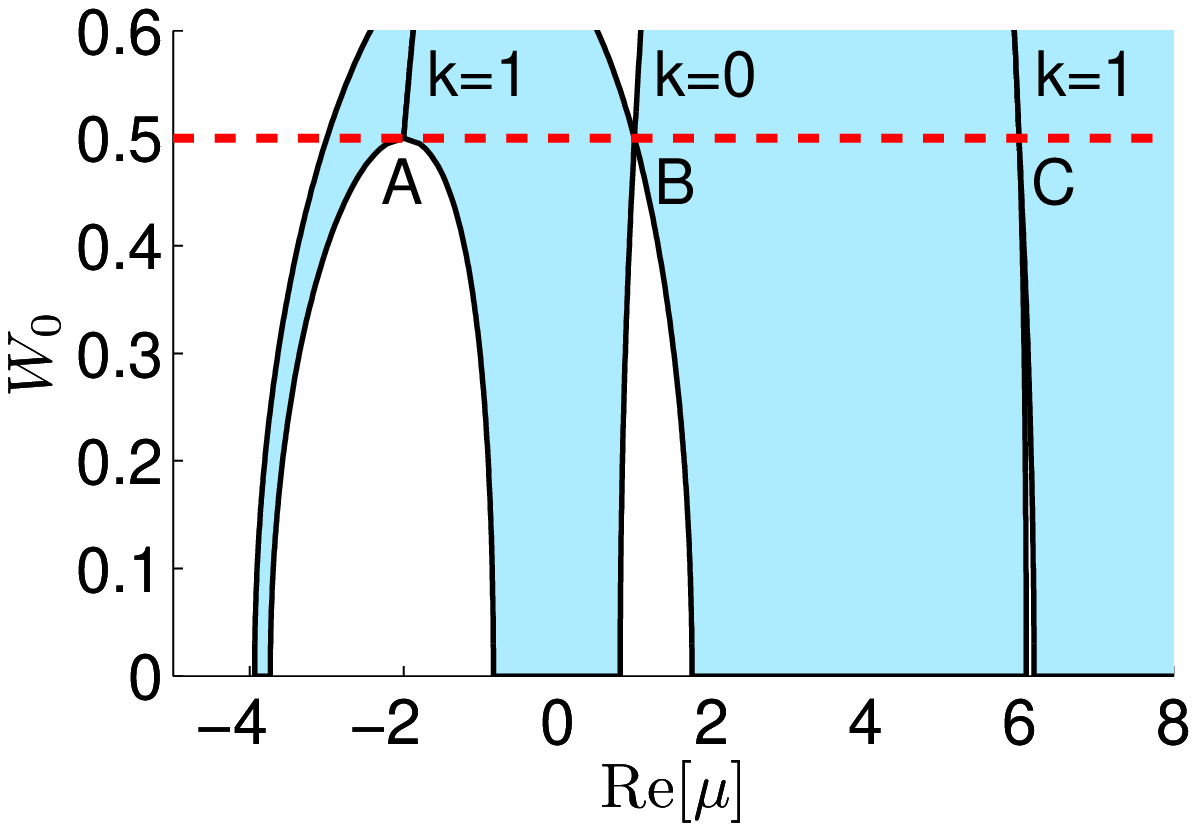}

\vspace{0.5cm}
\includegraphics[width=0.45\textwidth]{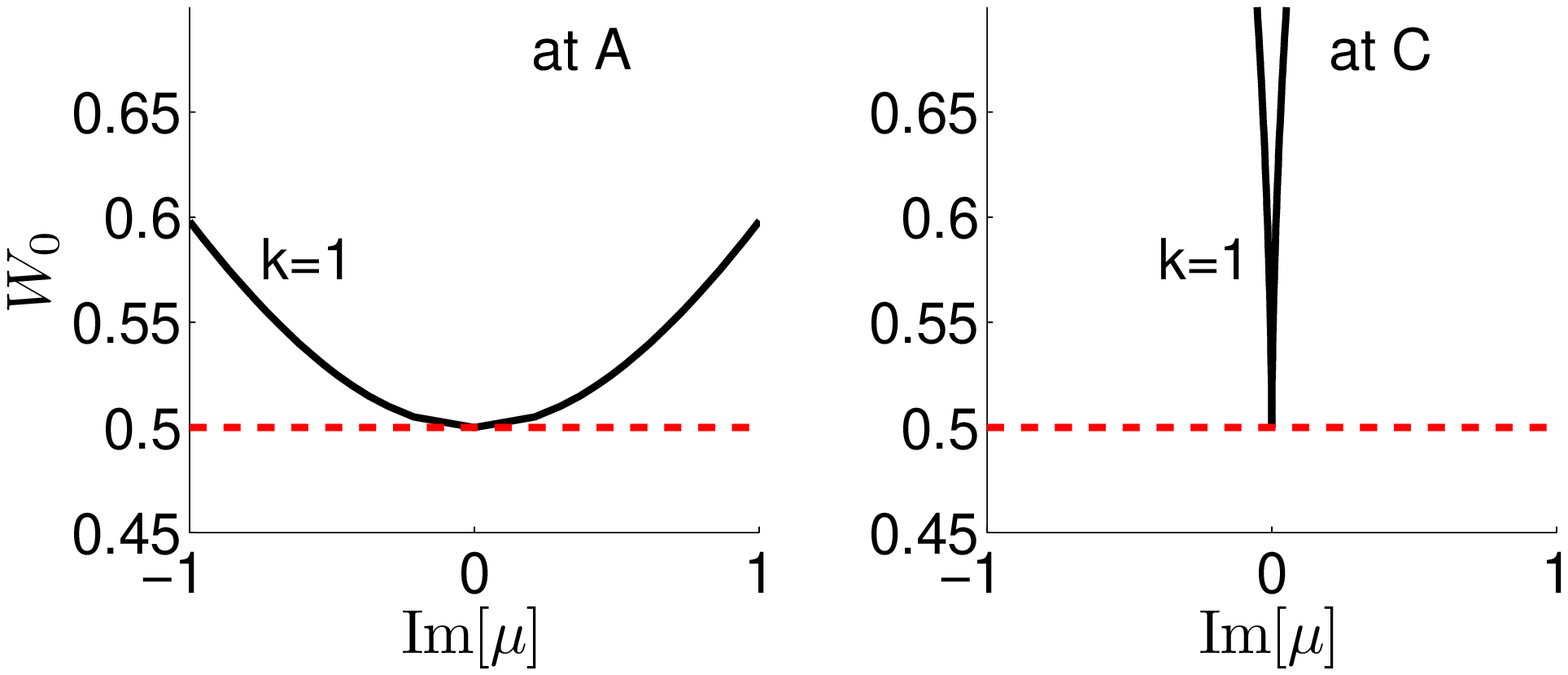}
\caption{(Color online) Bandgap structure of the PT lattice
(\ref{e:PTlattice}) as $W_0$ crosses the phase transition point
$0.5$ (with $V_0=6$). Above this phase transition point, complex
eigenvalues $\mu$ bifurcate out simultaneously from points A, C, ...
where Bloch bands merge (see the upper panel). The real and
imaginary parts of these complex eigenvalues versus $W_0$ at the
Brillouin edge $k=1$ are plotted in the upper and lower panels
respectively. } \label{f:fig2}
\end{figure}

For the PT lattice (\ref{e:PTlattice}), the phase transition point
is known to be $W_0 = 0.5$ \cite{Musslimani2008}. Below this phase
transition point ($W_0<0.5$), the continuous spectrum is all real
and comprises an infinite number of segments (known as Bloch bands).
The gaps between these Bloch bands are called bandgaps; the largest,
which contains everything to the left of the continuous spectrum, is
the semi-infinite gap and further gaps are numbered (in our case
from left to right). As an example, at $W_0=0.4$ and $V_0=6$, the
diffraction relation is shown in Fig. \ref{f:fig1} and the Bloch
bands and bandgaps are shown in Fig. \ref{f:fig2}.

As $W_0$ increases, bandgaps shrink (see Fig. \ref{f:fig2}). At the
phase transition point ($W_0 = 0.5$), all Bloch bands merge (see
Figs. \ref{f:fig1} and \ref{f:fig2}). Above the phase transition
point ($W_0>0.5$), complex eigenvalues appear in the Bloch bands.
This phase transition has been reported before
\cite{Musslimani2008}. For example, the diffraction relation at
$W_0=0.6$ and $V_0=6$ is displayed in Fig. \ref{f:fig1}. It is seen
that complex eigenvalues $\mu$ arise in the Bloch bands near edges
$k=\pm 1$ of the Brillouin zone.

What was not known about this phase transition, however, is that
right above this phase transition point, complex eigenvalues appear
simultaneously in an infinite number of Bloch bands. To demonstrate,
the dependence of eigenvalues $\mu$ on $W_0$ at Brillouin-zone edge
$k=1$ is shown in Fig. \ref{f:fig2}. We can see that at the phase
transition point $W_0 = 0.5$, complex eigenvalues $\mu$ bifurcate
out simultaneously from point A where the first and second Bloch
bands merge, and from point C where the third and fourth Bloch bands
merge, with both band-mergings occurring at the Brillouin-zone edges
$k=\pm 1$ (this bifurcation of complex eigenvalues does not occur
from point B where the second and third Bloch bands merge at the
Brillouin-zone center $k=0$).

Below, we show analytically that at the phase transition point, an
infinite number of complex eigenvalues bifurcate out simultaneously
from an infinite number of Bloch bands. In particular, bifurcations
of these complex eigenvalues occur at points where the $(2n-1)$-th
and the $2n$-th Bloch bands merge (at $k=\pm 1$), but not at points
where the $2n$-th and the $(2n+1)$-th Bloch bands merge (at $k=0$),
for all positive integers $n=1, 2, 3, \cdots$ (see Figs.
\ref{f:fig1} and \ref{f:fig2}).

We look for solutions to Eq. \eqref{e:linear1D} of the form $U =
u(x) \re^{-\ri \mu z}$, where $u$ satisfies the equation
 \begin{equation}
 \mu u + u_{xx} + V_0 \left( \cos^2x + \ri W_0 \sin 2x \right) u =
 0.
 \label{LinearMu}
 \end{equation}
At the phase transition point $W_0 = 0.5$, Eq. \eqref{LinearMu}
reduces to
  \begin{equation}
 \left(\mu + \frac{V_0}{2}\right) u + u_{xx} + \frac{V_0}{2} \left( \re^{2 \ri x} \right) u =
 0.
 \end{equation}
 Under the variable transformation $\xi= \ri \sqrt{V_0/2} \hspace{0.1cm} \re^{\ri
 x}$, this equation becomes Bessel's equation,
 \begin{equation}
 \xi^2 u_{\xi\xi}+\xi u_\xi +\left(\xi^2-\mu-\frac{V_0}{2}\right)u=0,
 \end{equation}
 thus it has exact solutions in terms of Bessel functions
 \begin{equation} \label{e:ubessel}
 u(x) = J_{ k }\left( \ri \sqrt{\frac{V_0}{2}} \hspace{0.1cm} \re^{\ri x}
 \right),
 \end{equation}
where $k = \pm\sqrt{\mu + \frac{V_0}{2}}$, or
\begin{equation} \label{e:exact_dispersion}
\mu = -\frac{V_0}{2} + k^2.
\end{equation}
This is the exact diffraction relation at the phase transition
point, as can be seen by utilizing the power-series expansion of the
Bessel function to expand the above Bloch solution (\ref{e:ubessel})
into a Fourier series
\begin{align}
J_{k}\left( \ri \sqrt{\frac{V_0}{2}} \re^{\ri x} \right) &=
\disp\sum_{m=0}^{\infty}
\frac{(V_0/8)^m}{m! \Gamma(m+ k + 1)}\re^{\ri(2 m + k)x} \nonumber \\
& \equiv  \re^{ik x} \rho(\re^{2\ri mx}),    \label{e:Jk}
\end{align}
where $k$ is seen to be the wavenumber and $\rho(\re^{2\ri mx})$ is
a $\pi$-periodic function. By factoring out the $\pi$-periodic term
$\re^{2inx}$ from $\re^{ik x}$ (for a certain integer $n$) and
combining it with $\rho(\re^{2\ri mx})$, one can restrict the
wavenumber $k$ to be in the Brillouin zone $-1\le k\le 1$, as is
customary in the Bloch theory (see Fig. \ref{f:fig1}). The
diffraction relation (\ref{e:exact_dispersion}) shows that the
continuous spectrum at the phase transition point $W_0=0.5$ is
$-V_0/2\le \mu <\infty$ and is entirely real. When $k=n$ is an
integer, the two Bessel solutions $J_{\pm k}(x)$ in
(\ref{e:ubessel}) are linearly dependent. This corresponds to the
points where different Bloch bands merge (see points A, B, C, ... in
Fig. \ref{f:fig2}), and the associated $\mu$ values are
\begin{equation}
\mu=-\frac{V_0}{2} + n^2, \quad n=0, 1, 2, \cdots.
\end{equation}
These $\mu$ values are located at either $k=0$ or $k=\pm 1$ of the
Brillouin zone on the diffraction curves, depending on whether $n$
is even or odd (see Fig. \ref{f:fig1}), and their Bloch functions
are $\pi$-periodic for even $n$ and $2\pi$-periodic for odd $n$.

We now consider the case where $W_0$ is near the phase transition
point 0.5, i.e., $V_0(W_0 - 0.5) \equiv \epsilon \ll 1$. In this
case, Eq. \eqref{LinearMu} becomes
\begin{equation}
 \left(\mu + \frac{V_0}{2}\right) u + u_{xx} + \frac{V_0}{2} \left( \re^{2 \ri x} \right) u + \epsilon \ri \sin (2x) =
 0,
 \label{PerturbedMu}
\end{equation}
whose solutions and the corresponding diffraction relation
$\mu=\mu(k)$ can be derived by the perturbation method. For
simplicity, we only derive its solutions $u(x)$ which are $\pi$- or
$2\pi$-periodic (these Bloch solutions are degenerate). The
corresponding $\mu$ values are then those with $k=0$ or $k=\pm 1$ on
the diffraction curves (see Fig. \ref{f:fig1}). These solutions and
the associated $\mu$ values can be expanded as power series in
$\epsilon^{1/2}$,
\begin{subequations}
\begin{align}
\mu &= - \frac{V_0}{2}  +  \rn_0^2 + \epsilon^{1/2} \rn_1 + \epsilon
\rn_2 + \epsilon^{3/2} \rn_3+ \ldots,   \label{muexpand}\\
u(x) & = u_0 +  \epsilon^{1/2} u_1 + \epsilon u_2 + \epsilon^{3/2}
u_1 + \ldots,
\end{align}
\label{Expansions}
\end{subequations}
where $\rn_0 = 0,1,2,\cdots$, and coefficients
$\rn_1,\rn_2,\rn_3,\cdots$ in (\ref{muexpand}) are certain
constants. Details of this perturbation calculation are presented in
Appendix 1. The main results for these coefficients
$\rn_1,\rn_2,\cdots$ at various $\rn_0$ values are summarized in the
following table.

\begin{table}[h]
\caption{\label{tab:fonts} Coefficients in the $\mu$ expansion
(\ref{muexpand}).}
\begin{center}
  \begin{tabular}{ | c | c | c|  r | }
    \hline
     $\rn_0$ & $\rn_1$ & $\rn_2$ & $\rn_3$ \\ \hline \hline
          0  &   0  & $ V_0/8 $& 0  \\ \hline
     1&   $ -\ri \disp\frac{V_0^{1/2}}{2}$, $ \ri \disp\frac{V_0^{1/2}}{2}$& $\disp V_0/32$ &   \\ \hline
     2 & 0 & $- \disp\frac{5V_0}{48}$, $\disp\frac{V_0}{48}$& 0 \\ \hline
     3 & 0 &  $-V_0/64$  & $-\ri \disp\frac{ ~V_0^{3/2}}{2^9}$,
     $\ri \disp\frac{ ~V_0^{3/2}}{2^9}$ \\ \hline
     N & 0 & $ - \disp\frac{V_0}{8} \disp\frac{1}{N^2 -1}$  & 0 \\ \hline
  \end{tabular}
\end{center}
\end{table}

We see from this table that when $\rn_0=1, 3$, which correspond to
points A, C in Fig. \ref{f:fig2}, the coefficient $\rn_1$ or $\rn_3$
is imaginary, thus complex eigenvalues bifurcate out simultaneously
above the phase transition point ($\epsilon>0$). In addition, the
imaginary part of these complex eigenvalues at $\rn_0=3$ is much
smaller than that at $\rn_0=1$ since the former is of order
$\epsilon^{3/2}$ while the latter is of order $\epsilon^{1/2}$.
However, no complex eigenvalues appear when $\rn_0=0, 2$ (the latter
corresponds to the point B in Fig. \ref{f:fig2}). All these
analytical results are in complete qualitative and quantitative
agreement with Fig. \ref{f:fig2} as we have carefully checked.
Continuing these calculations to higher $\rn_0$ values, we have
found that the coefficient $\rn_{2m+1}$ is always imaginary for
$\rn_0=2m+1$, where $m=0, 1, 2, \cdots$. Thus complex eigenvalues
bifurcate out simultaneously from all odd values of $\rn_0$ at the
phase transition point $W_0=0.5$.

The above table also shows that below the phase transition point
($W_0<0.5$, or $\epsilon<0$), the eigenvalue $\mu$ from the
expansion (\ref{muexpand}) is always real for all integers $\rn_0$.
In addition, a gap opens at the corresponding $\mu$ values of
$-V_0/2+\rn_0^2$. Furthermore, the width of the $\rn^{\rm th}$ gap
is of order $\epsilon^{\rn/2}$. Above the phase transition point,
the even-numbered bandgaps reopen, whereas the odd-numbered bandgaps
close and complex eigenvalues bifurcate out. All these analytical
conclusions match perfectly with Fig. \ref{f:fig2} as well.

Now we consider eigenvalue bifurcations in two dimensions. In this
case, the 2D linear Schr\"odinger equation (\ref{e:model}) is
\begin{equation} \label{e:linear2D}
i U_z + U_{xx} +U_{yy} + V(x, y)U = 0,
\end{equation}
where the PT lattice potential $V(x,y)$ is given in Eq.
(\ref{e:PTlattice2D}). This 2D potential is separable, thus the
Bloch modes of Eq. (\ref{e:linear2D}) are \cite{Yang_book}
\begin{equation}
U(x,y,z)=e^{ik_1x+ik_2y-i\mu z}p(x; k_1)p(y; k_2),
\end{equation}
where $p(x; k)$ is the 1D $\pi$-periodic function as given in
(\ref{e:Bloch1D}),
\begin{equation} \label{e:2Ddiff}
\mu=\widehat{\mu}(k_1)+\widehat{\mu}(k_2)
\end{equation}
is the 2D diffraction relation, $k_1, k_2$ are Bloch wavenumbers in
$x$ and $y$ directions and are located inside the irreducible
Brillouin zone $-1\le k_1, k_2 \le 1$, and the function
$\widehat{\mu}(k)$ is the diffraction relation of the 1D equation
(\ref{e:linear1D}). This diffraction relation (\ref{e:2Ddiff}) shows
that complex eigenvalues appear in this 2D PT lattice if and only if
complex eigenvalues appear in the 1D PT lattice (\ref{e:PTlattice}).
Thus all eigenvalues in the 2D system (\ref{e:linear2D}) are real
when $W_0\le 0.5$, and a phase transition occurs at $W_0=0.5$ above
which complex eigenvalues arise. In addition, an infinite number of
Bloch bands turn complex simultaneously right above this phase
transition point.

\section{Stability of PT solitons in one dimension}
\label{Sec: 1DStability}

In the presence of cubic nonlinearity, the mathematical model
becomes the NLS equation (\ref{e:model}) with a PT lattice
potential. In this case, light can self-localize and form solitons.
In this section, we study these PT solitons and their
linear-stability behaviors in one dimension.

In one dimension, the NLS equation (\ref{e:model}) becomes
\begin{equation}
\ri U_z + U_{xx} + V(x)U + \sigma |U|^2U = 0.  \label{PTNLS}
\end{equation}
Here the PT lattice $V(x)$ is taken as (\ref{e:PTlattice}) with
$V_0=6$, and $\sigma=\pm 1$. Solitons in this model are sought of
the form
\begin{equation}
U(x,z)=\re^{-\ri\mu z}u(x),
\end{equation}
where $u(x)$ is a localized function, and $\mu$ is a real
propagation constant. These solitons can be computed by either the
squared operator iteration method or the Newton-conjugate-gradient
method applied to the normal equation \cite{Yang_book}. They exist
when $\mu$ lies inside bandgaps of the linear system for $W_0$ both
below and above the phase transition point. Above the phase
transition point ($W_0>0.5$), linear waves amplify exponentially
during propagation, thus any solitons would also be unstable to
perturbations. So we only need to consider $W_0\le 0.5$ below.

To determine the linear stability of these PT solitons, we perturb
them as
\begin{equation}
U = \re^{-\ri \mu z} \left[ u(x) + \tilde{u}(x) \hspace{0.05cm}
\re^{ \lambda z} + \tilde{w}^*(x) \hspace{0.05cm} \re^{ \lambda^* z}
\right],
\end{equation}
where $|\tilde{u}|,  |\tilde{w}|\ll |u|$. After substitution into
equation \eqref{PTNLS} and linearizing, we arrive at the eigenvalue
problem
\begin{equation}
\ri {\cal L} \left(\begin{array}{c} \tilde{u}\\ \tilde{w}
\end{array} \right) = \lambda   \left(\begin{array}{c} \tilde{u}\\ \tilde{w}
\end{array}\right),  \label{EigenProblem}
\end{equation}
where
\begin{align*}
{\cal L} &= \left( \begin{array}{c c} L_{11} &  L_{12} \\
L_{21} & L_{22}\end{array} \right), \\
L_{11} &= \mu +  \partial_{xx}  +  V(x)  + 2 \sigma |u|^2,  \\
L_{12} &= \sigma u^2, \\
L_{21} &= -\sigma \left(u^2\right)^*,  \\
L_{22} &= -\left(\mu +  \partial_{xx}  +  V^*(x)  + 2\sigma |u|^2
\right).
\end{align*}
This eigenvalue problem can be computed by the Fourier collocation
method (for the full spectrum) or the Newton-conjugate-gradient
method (for individual discrete eigenvalues) \cite{Yang_book}. If
eigenvalues with positive real parts exist, the soliton is linearly
unstable; otherwise it is linearly stable.

\begin{figure}[!htbp]
\centering
\includegraphics[width=0.45\textwidth]{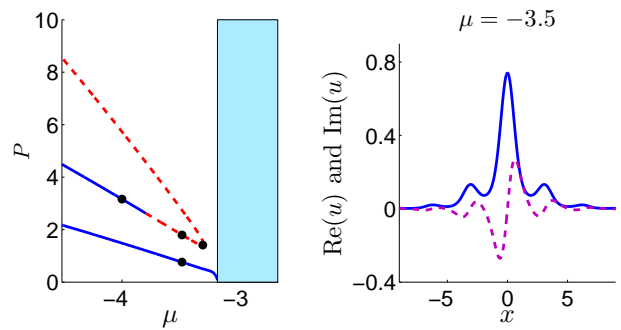}
\caption{(Color online) One-dimensional PT solitons in the
semi-infinite gap under focusing nonlinearity ($\sigma=1$) for
$V_0=6$ and $W_0=0.45$. (left) Power curves of these solitons; the
lower curve is for fundamental solitons and the upper curve for
dipole solitons; solid and dashed lines represent stable and
unstable solitons respectively (the same holds for all other
figures); the shaded region is the first Bloch band. (right) Profile
$u(x)$ of a fundamental soliton at $\mu = -3.5$ (marked by a dot on
the lower curve of the left panel); the solid blue line is for the
real part and dashed pink line for the imaginary part. }
\label{PowerCurve1DFocusing}
\end{figure}

\begin{figure}[!htbp]
\centering
\includegraphics[width=0.5\textwidth]{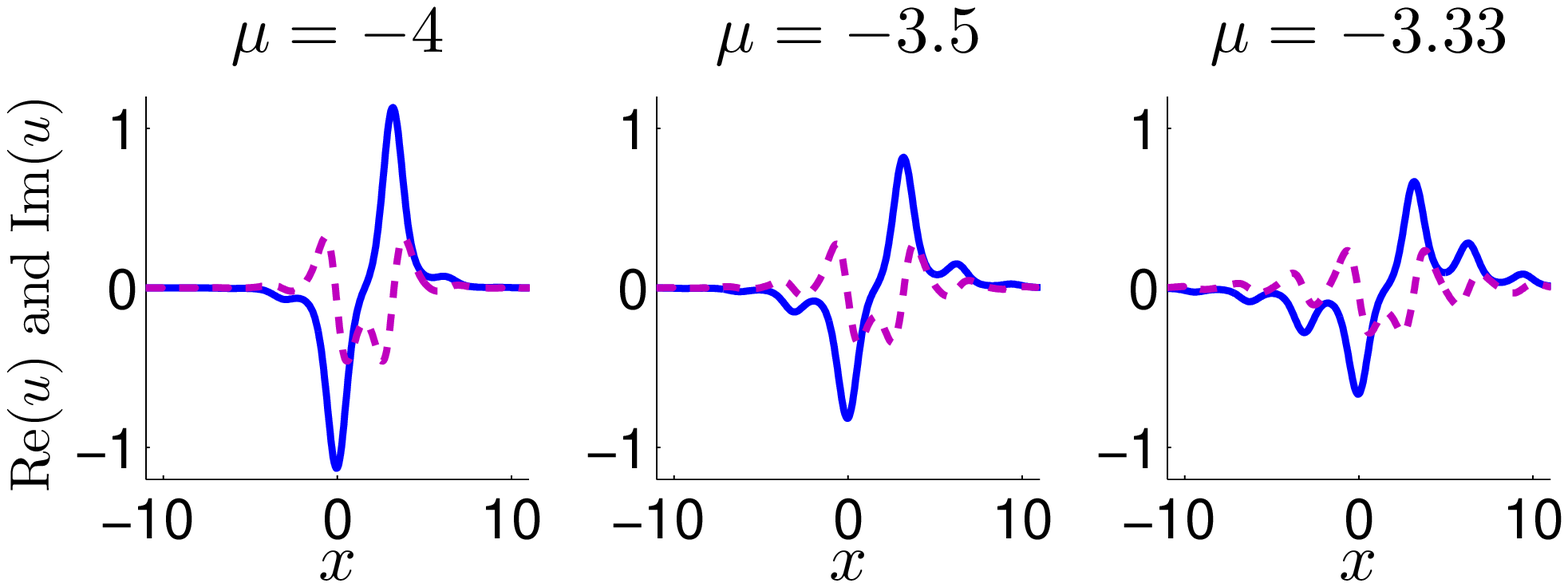}

\vspace{0.5cm}
\includegraphics[width=0.5\textwidth]{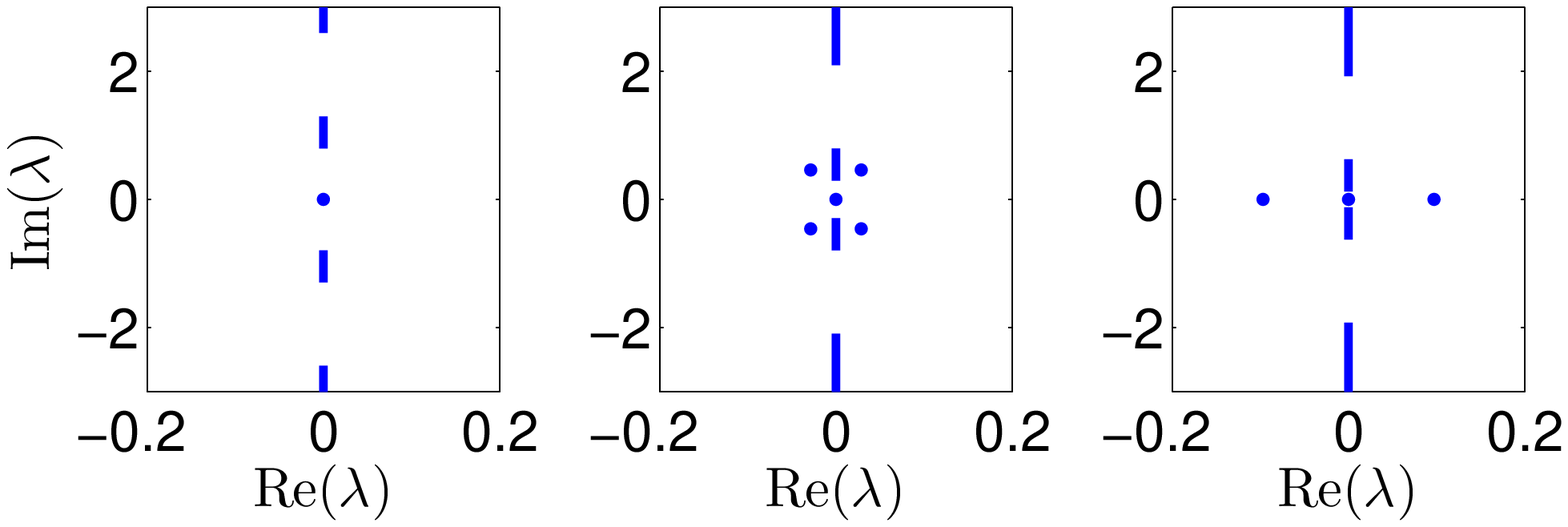}
\caption{(Color online) Dipole solitons (top) and their
linear-stability spectra (bottom) for three $\mu$ values in the
semi-infinite gap. The power curve of these dipole solitons is shown
in Fig. \ref{PowerCurve1DFocusing}, and the locations of these
solitons are marked by dots on that power curve. }
\label{Spectrum1DFocusing}
\end{figure}

We first consider PT solitons in the semi-infinite gap under
focusing nonlinearity ($\sigma = 1$). For $W_0=0.45$, two families
of PT solitons are obtained and their power curves are displayed in
Fig. \ref{PowerCurve1DFocusing} (left). Here the power of a soliton
is defined as
\begin{equation}
P(\mu)=\int_{-\infty}^\infty |u(x; \mu)|^2 \rd x.
\end{equation}
In this figure, the lower power curve is for the fundamental
solitons which exhibit the same PT symmetry $u^*(x)=u(-x)$ and whose
real parts possess a single dominant peak. The profile of such a
soliton at $\mu=-3.5$ is displayed in Fig.
\ref{PowerCurve1DFocusing} (right). This soliton family bifurcates
out of the first Bloch band, and the solitons near this Bloch band
are low-amplitude Bloch-wave packets. We have found that the entire
branch of this fundamental-soliton family is linearly stable, which
is indicated by solid lines of its power curve in Fig.
\ref{PowerCurve1DFocusing} (left). The upper power curve in Fig.
\ref{PowerCurve1DFocusing} is for the dipole solitons. This power
curve features double branches which terminate before reaching the
first Bloch band (a similar phenomenon occurs in purely real
lattices \cite{Yang_book,Akylas}). Profiles of three such solitons
on the lower power branch are displayed in Fig.
\ref{Spectrum1DFocusing} (top). It is seen that the real parts of
these dipole solitons possess two dominant peaks of opposite phase
(which is why they are termed dipole solitons). Unlike the
fundamental solitons, these dipole solitons are linearly stable only
in a certain portion of their existence region. Specifically, only
dipole solitons on the lower branch and with $\mu < -3.8$ are stable
(see Fig. \ref{PowerCurve1DFocusing} (left)). For dipole solitons in
this region, their spectra are entirely imaginary (see Fig.
\ref{Spectrum1DFocusing} (bottom left)). At $\mu = -3.8$, stability
switching occurs where a quadruple of complex eigenvalues bifurcate
off of the edge of the continuous spectrum (see Fig.
\ref{Spectrum1DFocusing} (bottom center)). Within the unstable
region, there is a second eigenvalue bifurcation at $\mu\approx
-3.4$ of the lower branch (near and on the left side of the power
minimum) where a pair of real eigenvalues bifurcate from zero (see
Fig. \ref{Spectrum1DFocusing} (bottom right)). Some of these
stability behaviors on dipole solitons are similar to those in the
purely real potential ($W_0 = 0$) \cite{Yang_book}. A notable
difference is that for real potentials real eigenvalues bifurcate
out of the origin exactly at the minimum of the power curve
\cite{Yang_book}, whereas here this real-eigenvalue bifurcation
occurs \emph{not} at the power minimum. An analytical explanation
for this new phenomanon will be given in Appendix 2.

\begin{figure}[!htbp]
\centering
\includegraphics[width=0.45\textwidth]{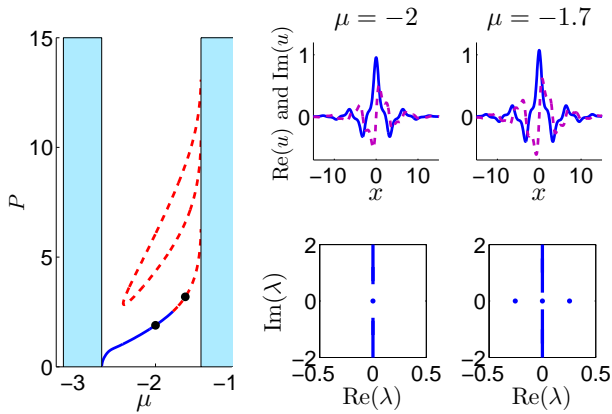}
\caption{ (Color online) One-dimensional PT solitons in the first
gap under defocusing nonlinearity ($\sigma=-1$) for $V_0=6$ and
$W_0=0.45$. (left) Power curves of these solitons; the lower curve
is for fundamental solitons and the upper curve for dipole solitons;
(top right) two fundamental solitons at $\mu=-2$ and $-1.7$ (marked
by dots in the left panel); (bottom right) linear-stability spectra
of these solitons. } \label{PowerCurve1DDefocusing}
\end{figure}

\begin{figure}[!htbp]
\centering
\includegraphics[width=0.45\textwidth]{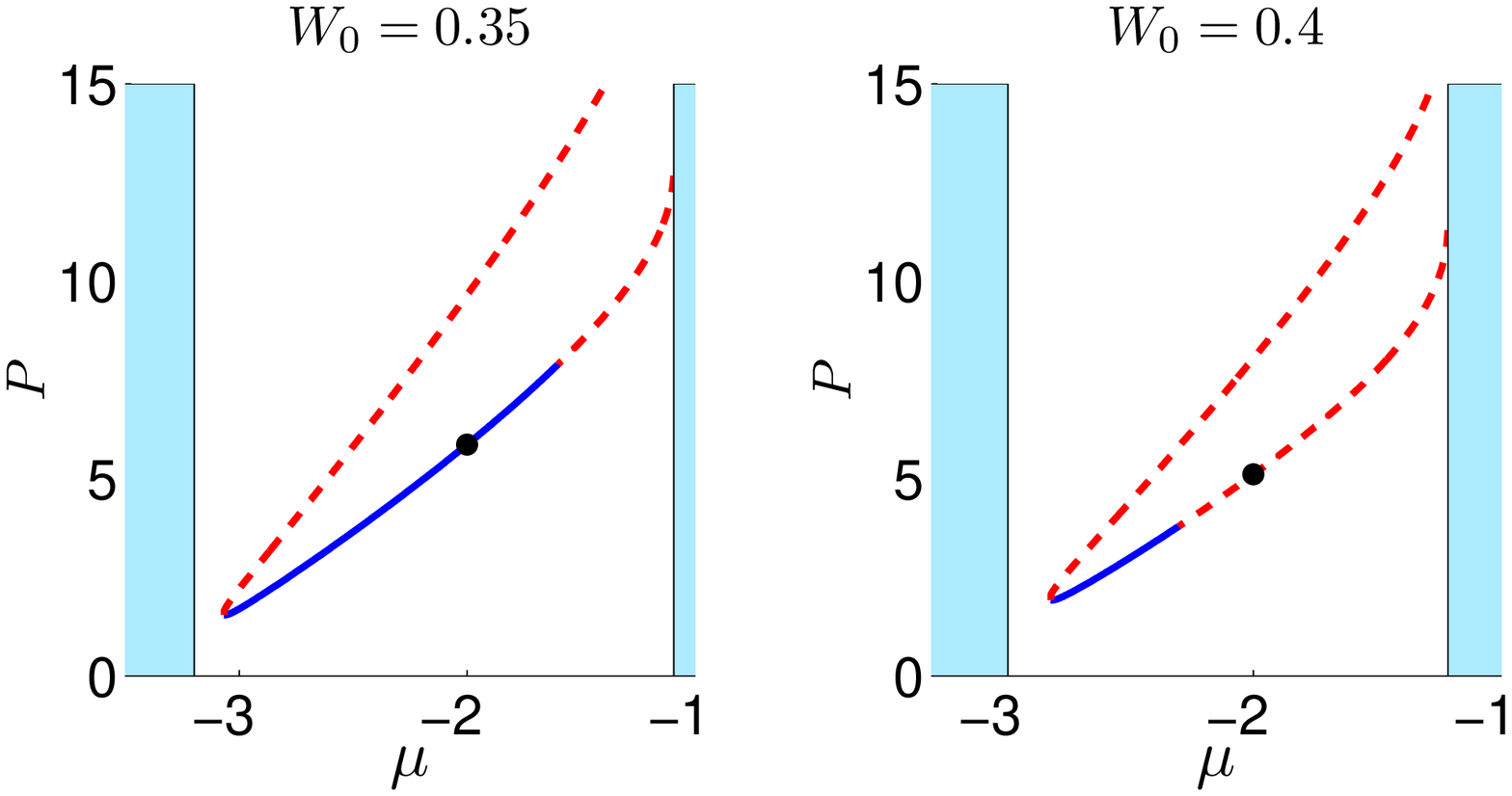}

\vspace{0.3cm}
\includegraphics[width=0.45\textwidth]{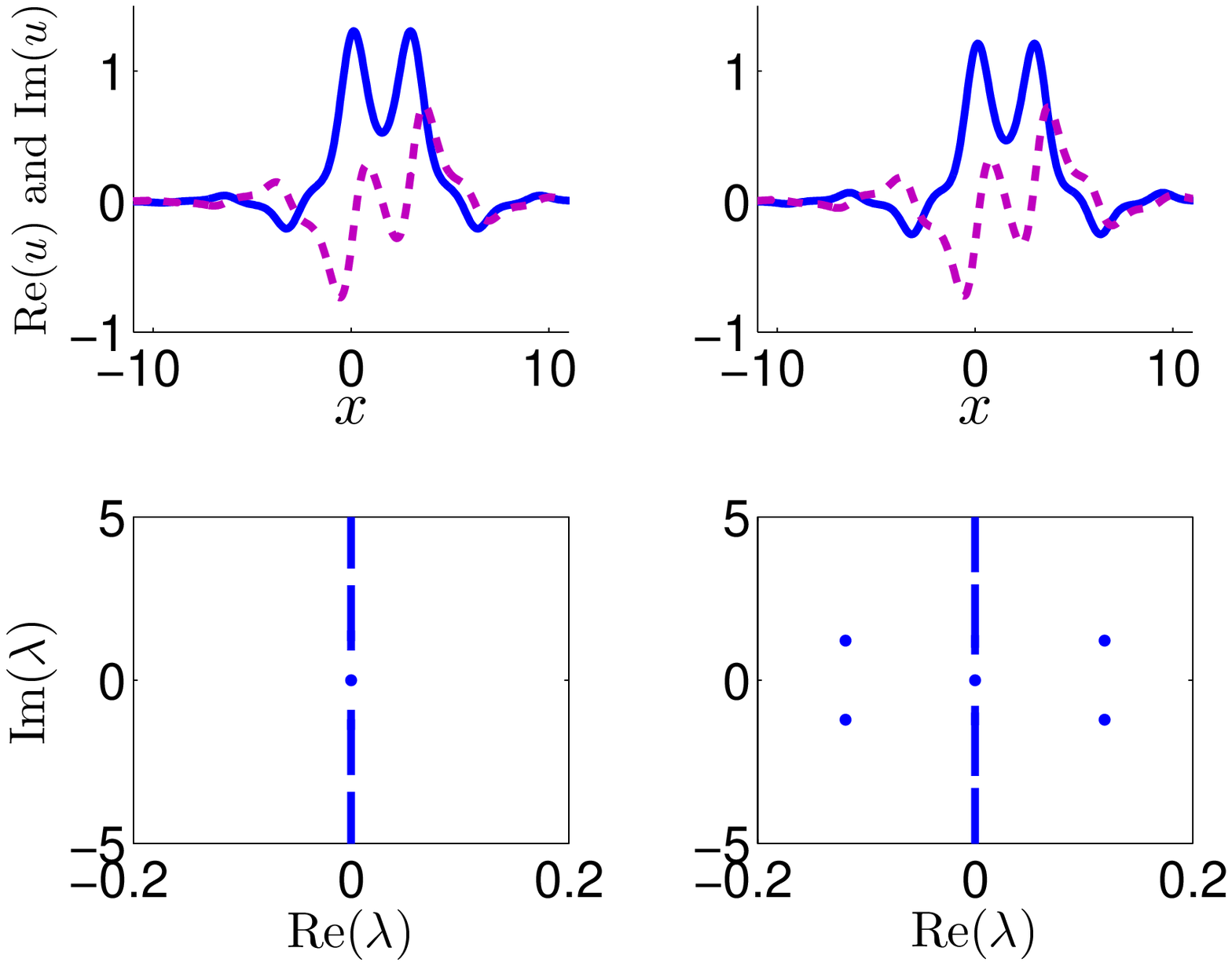}
\caption{(Color online) Dipole solitons in the first gap under
defocusing nonlinearity ($\sigma=-1$) at $W_0$ values 0.35 and 0.4
(with $V_0=6$). (top) Power curves; (middle) soliton profiles at
$\mu=-2$ (marked by dots in the top panel); (bottom)
linear-stability spectra of the solitons in the middle panel.}
\label{Spectrum1DDefocusing2}
\end{figure}

Next we consider PT solitons in the first gap under defocusing
nonlinearity ($\sigma = -1$). Again, for $W_0=0.45$, two families of
PT solitons are obtained and their power curves are displayed in
Fig. \ref{PowerCurve1DDefocusing} (left) with stability results
indicated. The lower curve is for fundamental solitons whose
profiles at two $\mu$ values are depicted in Fig.
\ref{PowerCurve1DDefocusing} (top right panel), while the upper
curve is for dipole solitons, whose profiles are similar to those in
Fig. \ref{Spectrum1DDefocusing2} (middle panel) below. The
fundamental-soliton family bifurcates out of the first Bloch band,
whereas the dipole family does not. We have found that all solitons
in this dipole family are linearly unstable (see Fig.
\ref{PowerCurve1DDefocusing} (left)). The fundamental-soliton
family, however, is linearly stable when $\mu< -1.77$. At
$\mu=-1.77$, stability switching occurs where a pair of real
eigenvalues bifurcate out from zero (see Fig.
\ref{PowerCurve1DDefocusing}). Notice that unlike in real potentials
\cite{Yang_book}, this zero-eigenvalue bifurcation does not occur at
a power extremum since the potential here is complex. An explanation
for this will be presented in Appendix 2.

The stability results of PT solitons in Figs.
\ref{PowerCurve1DFocusing} to \ref{PowerCurve1DDefocusing} were
obtained for a specific $W_0$ value of 0.45. Now we discuss how
these stability results change when $W_0$ steadily increases from 0
to 0.5. First we consider PT solitons in the semi-infinite gap under
focusing nonlinearity. We find that when $0\le W_0< 0.5$, the entire
fundamental-soliton family remains stable. The dipole family,
however, is stable only on the left side of its lower branch (see
Fig. \ref{PowerCurve1DFocusing}), and this stable region shrinks as
$W_0$ increases. Next we consider PT solitons in the first gap under
defocusing nonlinearity. When $W_0=0$ (i.e., the lattice is real),
the fundamental-soliton family is stable in the entire first gap
\cite{Yang_book}. As $W_0$ rises above 0.3, an unstable region grows
off the edge of the second band. At the stability switching point a
pair of real eigenvalues bifurcate from zero as illustrated in Fig.
\ref{PowerCurve1DDefocusing}. Regarding the dipole-soliton family,
its entire upper branch is unstable for all $W_0$ values. Its entire
lower branch is also unstable when $W_0>0.44$. For $W_0<0.44$, a
certain portion of its lower branch is stable; but as $W_0$
increases, this stable region shrinks and then totally disappears
when $W_0>0.44$. To demonstrate this reduced stability of dipole
solitons with increasing $W_0$, the power curves of these dipole
solitons at two $W_0$ values of 0.35 and 0.4 are shown in Fig.
\ref{Spectrum1DDefocusing2} (top panel) with stability results
indicated. The soliton profiles at $\mu=-2$ are also shown in the
middle panel of the same figure. It is seen that the stable region
of dipole solitons at $W_0=0.4$ is much shorter than that at
$W_0=0.35$. Notice also that as $W_0$ increases the width of the
first gap decreases which is often a sign of decreased stability.
The unstable region on the lower branch is largely located near the
edge of the second Bloch band, and the instability in this region is
caused by a quadruple of complex eigenvalues (see Fig.
\ref{Spectrum1DDefocusing2} (bottom panel)).

The above stability results of PT solitons show that as $W_0$
increases (but still below the phase transition point), the stable
regions of PT solitons generally shrink (see Fig.
\ref{Spectrum1DDefocusing2}). The only exception is the
fundamental-soliton family in the semi-infinite gap under focusing
nonlinearity, which remains entirely stable up to the phase
transition point. Overall, the inclusion of the gain-loss component
in the PT lattice has a destabilizing effect on solitons.

\section{Stability of PT solitons in two dimensions}
\label{Sec: 2DStability}

In this section we analyze the linear stability of solitons in a
two-dimensional PT lattice. We will show that the destabilizing
effect of the gain-loss component is more prominent in this case,
even for fundamental solitons in the semi-infinite gap.

In two dimensions, the mathematical model is Eq. (\ref{e:model}), or
\begin{equation} \label{e:model2}
i U_z + U_{xx} + U_{yy} + V(x,y)U + \sigma |U|^2U = 0,
\end{equation}
where the PT lattice $V(x,y)$ is taken as (\ref{e:PTlattice2D}) with
$V_0=6$. Solitons in this model are sought of the form
\begin{equation}
U(x,y,z)=\re^{-\ri\mu z}u(x,y),
\end{equation}
where $u(x,y)$ is a localized function, and $\mu$ is a real
propagation constant. These solitons as well as their
linear-stability spectra can be obtained by numerical methods
similar to the 1D case. The phase transition point in this 2D model
is also $W_0=0.5$, above which all solitons are linearly unstable.
Thus we only consider $W_0<0.5$ below.

\begin{figure}[!htbp]
\centering
\includegraphics[width=0.45\textwidth]{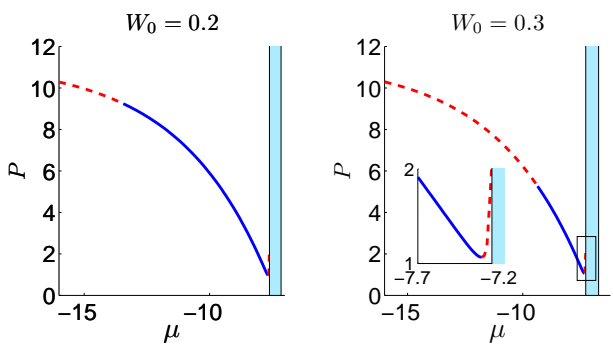}
\caption{(Color online) Power curves of fundamental 2D solitons in
the semi-infinite gap under focusing nonlinearity ($\sigma=1$) for
$V_0=6$ and two $W_0$ values of 0.2 and 0.3. The inset in the right
panel is amplification of the power curve near the first Bloch band
in the same panel. } \label{PowerCurve2D}
\end{figure}

\begin{figure}[!htbp]
\centering
\includegraphics[width=0.45\textwidth]{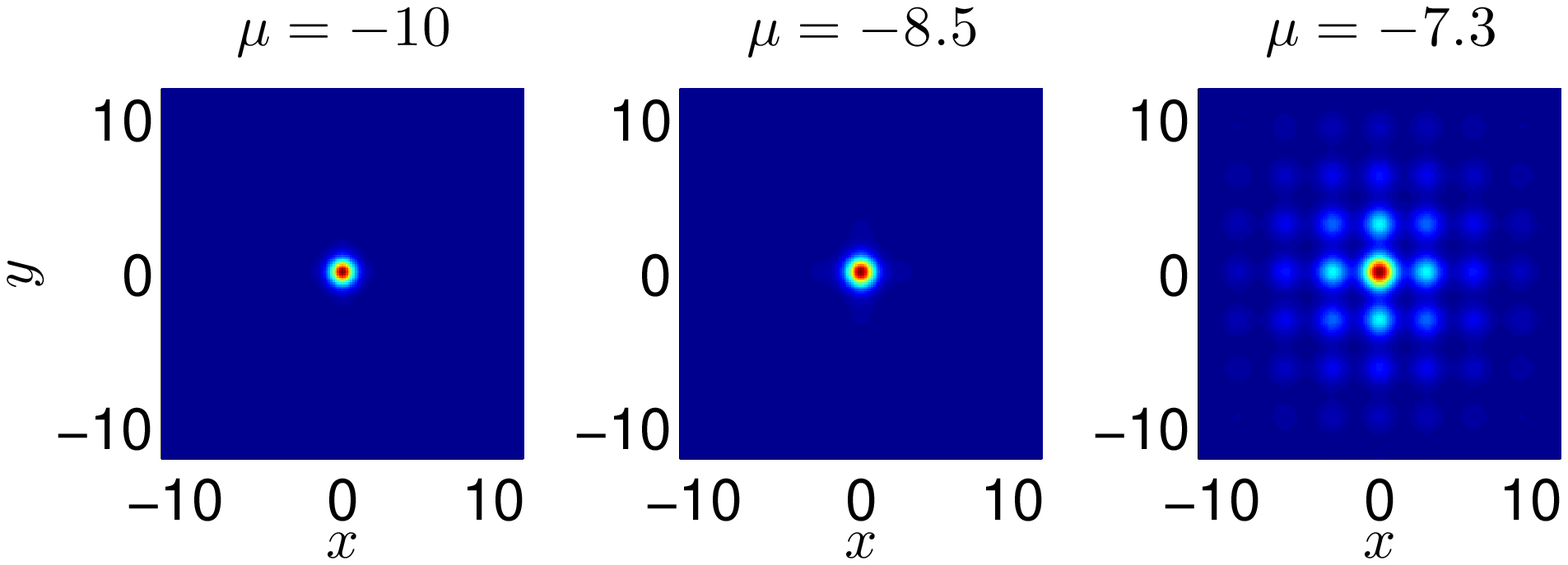}

\vspace{0.5cm}
\includegraphics[width=0.45\textwidth]{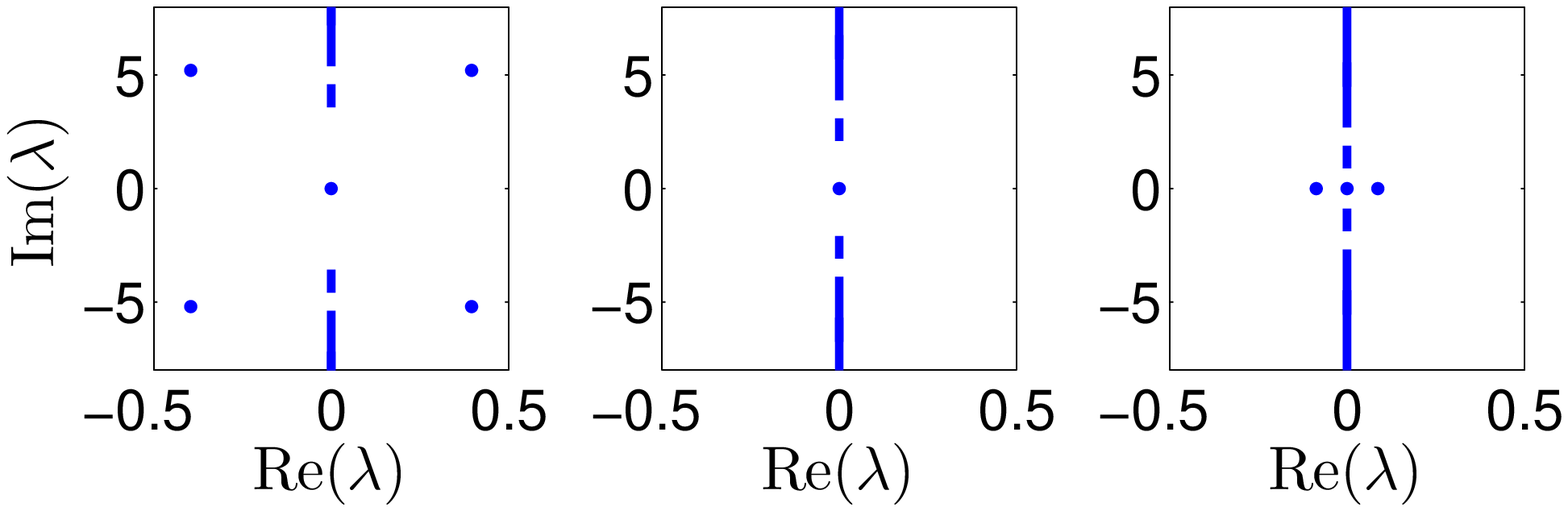}
\caption{(Color online) Fundamental 2D solitons ($|u(x,y)|$) (top)
and their linear-stability spectra (bottom) for three $\mu$ values
in the semi-infinite gap with $\sigma=1$, $V_0=6$ and $W_0=0.3$. The
power curve of these solitons is shown in Fig. \ref{PowerCurve2D}
(right panel). } \label{Spectrum2D}
\end{figure}

For simplicity we only consider 2D fundamental PT solitons in the
semi-infinite gap under focusing nonlinearity ($\sigma=1$). These
fundamental solitons possess the PT symmetry $u*(x,y)=u(-x,-y)$, and
their real parts have a single dominant peak. Profiles of such
solitons can be found in Fig. \ref{Spectrum2D} (upper panel) later.
We find that these fundamental solitons are stable only on a finite
$\mu$-interval even for small values of $W_0$. In addition, this
stable region shrinks as $W_0$ increases and totally disappears when
$W_0>0.47$. To demonstrate, power curves of these solitons as well
as their stability regions at two $W_0$ values of 0.2 and 0.3 are
displayed in Fig. \ref{PowerCurve2D}. It is seen that the stable
region is finite even though the existence region of solitons is
infinite. In addition, as $W_0$ increases from 0.2 to 0.3, the
stable region has shortened by several times. For each $W_0$, there
are two unstable regions, one located at large negative $\mu$
values, and the other one located near the first Bloch band. For
large negative values of $\mu$ the instability is due to a quadruple
of complex eigenvalues, whereas for $\mu$ values near the first
band, the instability is due to a pair of real eigenvalues. Examples
of the spectrum in each region are shown in Fig. \ref{Spectrum2D}
with $W_0=0.3$. We see that in this 2D case, even the fundamental
solitons in the semi-infinite gap are destabilized by the addition
of the gain-loss component in the lattice.

\section{Nonlinear evolution of PT solitons under perturbations}

In this section, we examine the nonlinear evolution of PT solitons
under weak perturbations. We find that when a PT soliton is linearly
stable, then it is also nonlinearly stable and propagates robustly
against perturbations. If the soliton is linearly unstable, then it
breaks up under perturbations, and its amplitude and energy can grow
unbounded over distance.

\begin{figure}[!htbp]
\centering
\includegraphics[width=0.45\textwidth]{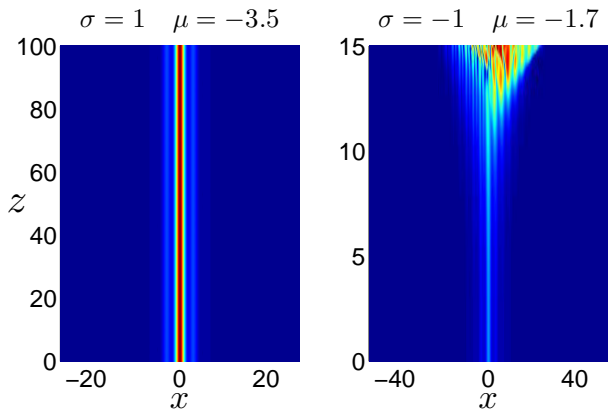}
\caption{(Color online) (left) Nonlinear evolution of the stable 1D
soliton in Fig. \ref{PowerCurve1DFocusing} under 5\% random noise
perturbations;  (right) Nonlinear evolution of the unstable 1D
soliton in Fig. \ref{PowerCurve1DDefocusing} (with $\mu=-1.7$) under
5\% random noise perturbations. Shown is the field $|U(x,z)|$ in the
$(x,z)$ plane. } \label{f:evolution1D}
\end{figure}

First we consider the 1D fundamental soliton shown in Fig.
\ref{PowerCurve1DFocusing}, which resides in the semi-infinite gap
under focusing nonlinearity and is linearly stable. We perturb it by
5\% random noise perturbations and then simulate its evolution in
Eq. (\ref{PTNLS}). The simulation result is shown in Fig.
\ref{f:evolution1D} (left). We can see that even after $z=100$ units
of propagation, this soliton remains robust and does not break up.
Thus this soliton is also nonlinearly stable. Next we consider the
1D fundamental soliton shown in Fig. \ref{PowerCurve1DDefocusing},
which resides in the first gap under defocusing nonlinearity and is
linearly unstable. When this soliton is perturbed by 5\% random
noise perturbations, its evolution is shown in Fig.
\ref{f:evolution1D} (right). It is seen that this soliton quickly
blows up and spreads out, thus is obviously nonlinearly unstable.
Notice that the peak amplitude and energy of this perturbed soliton
steadily increase without bound over distance. This indicates that
the gain-loss component of the PT lattice steadily feeds energy into
the solution. Recall that the $W_0$ value in this case is below the
phase transition point, thus linear waves do not grow. Consequently
the energy growth in this evolution is solely due to the nonlinear
effects.

\begin{figure}[!htbp]
\centering
\includegraphics[width=0.2\textwidth]{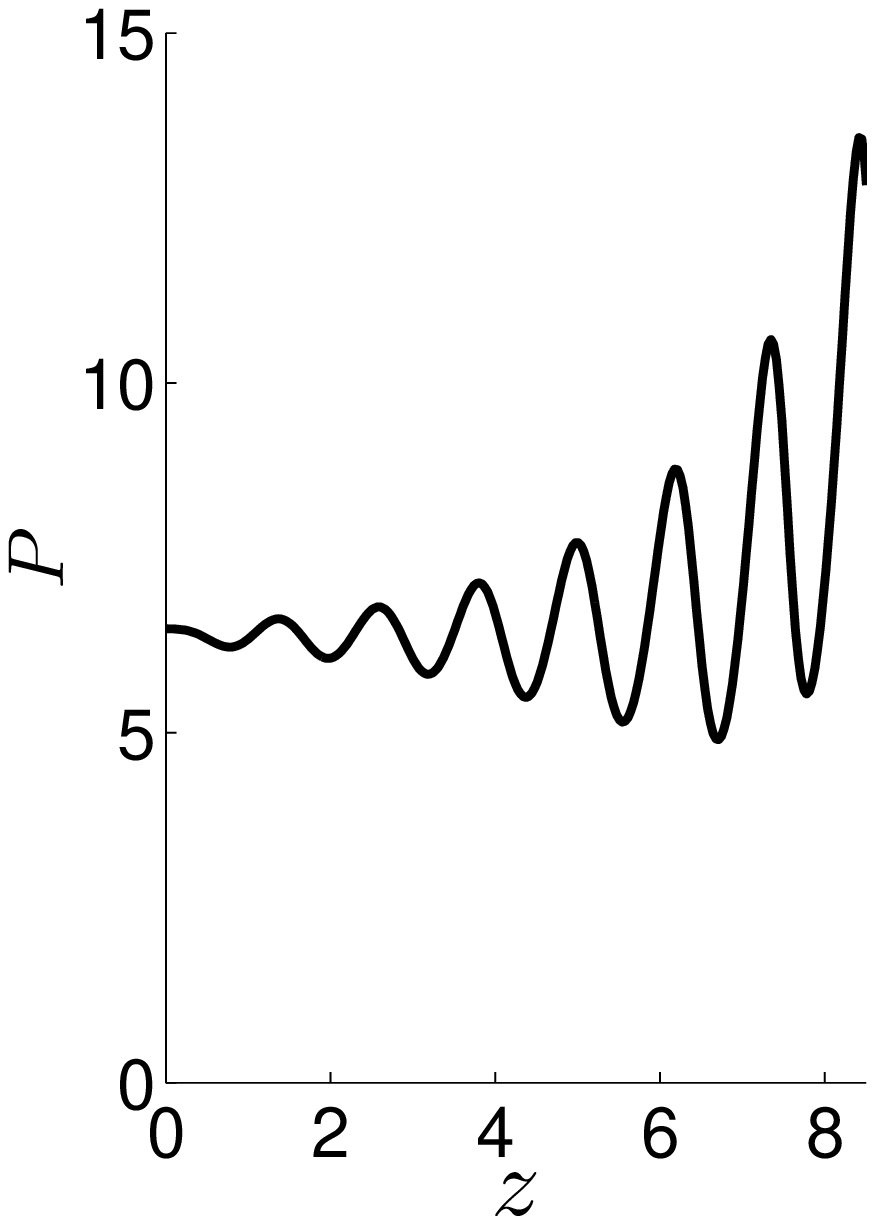}
\hspace{0.3cm}
\includegraphics[width=0.2\textwidth]{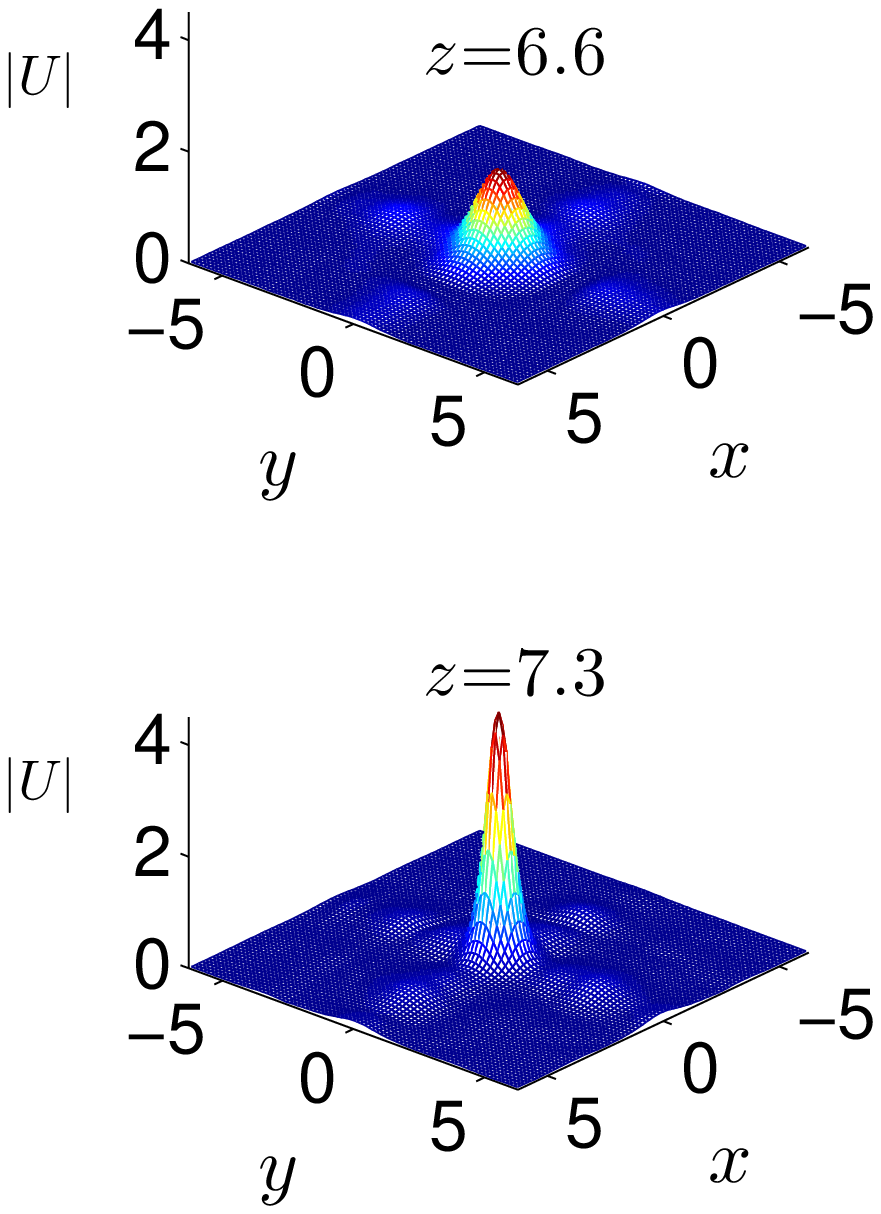}
\caption{(Color online) Nonlinear evolution of the unstable 2D
soliton in Fig. \ref{Spectrum2D} (with $\mu=-10$) under 5\% random
noise perturbations. (left) Power evolution versus distance $z$;
(right) Solution profiles at two distances $z=6.6$ and 7.3. }
\label{f:evolution2D}
\end{figure}

Lastly we consider the 2D fundamental soliton shown in Fig.
\ref{Spectrum2D} (left panel), which resides in the semi-infinite
gap under focusing nonlinearity and is linearly unstable. When this
soliton is perturbed by 5\% random noise perturbations, its
evolution is shown in Fig. \ref{f:evolution2D}. It is seen that the
power (and peak amplitude) of this perturbed soliton also grows
oscillatorily without bound, thus this soliton is nonlinearly
unstable. This oscillatory growth occurs since the unstable
eigenvalues of this soliton are complex (see Fig. \ref{Spectrum2D}
(lower left panel)).

\section{Summary}

In summary, we have analyzed the linear phase transition and
nonlinear solitons in PT-symmetric photonic lattices. We have shown
that at the phase transition point, an infinite number of linear
Bloch bands turn complex simultaneously. We have also shown that
while continuous ranges of stable solitons can exist in PT lattices,
increasing the gain-loss component of the lattice has an overall
destabilizing effect on soliton propagation. In addition, we have
shown that when unstable PT solitons are perturbed, the energy of
the solution can grow unbounded even though the PT lattice is below
the phase transition point.

\section*{Acknowledgment} This work of S.N. and J.Y. is supported in part by the Air Force
Office of Scientific Research (Grant USAF 9550-09-1-0228) and the
National Science Foundation (Grant DMS-0908167). The work of L.G. is
supported by a Visiting Student Scholarship from the Chinese
Scholarship Council.

\section*{Appendix 1: Calculation of eigenvalue bifurcations at the phase transition point}
\renewcommand{\theequation}{A.\arabic{equation}}

In this appendix, we calculate eigenvalue bifurcations at the phase
transition point in Eq. (\ref{PerturbedMu}) by perturbation methods.
The solution $u(x)$ to this equation is required to be $\pi$- or
$2\pi$-periodic, and perturbation expansions for $u(x)$ and
eigenvalue $\mu$ are as given in Eq. (\ref{Expansions}).

Let us define the operator
\begin{equation}
L = \partial_{xx} + \rn_0^2 +  \frac{V_0}{2} \re^{2 \ri x}.
\end{equation}
After substituting expansions \eqref{Expansions} into Eq.
\eqref{PerturbedMu} and collecting terms of the same order in
$\epsilon^{1/2}$ we arrive at the following system of linear
equations
\begin{subequations}
\begin{align}
L u_0  & = 0, \label{order0}\\
L u_1   & = -\rn_1 u_0,  \label{order1}\\
L u_m  & =  -\ri \sin(2x) u_{m-2} - \disp\sum_{j=1}^m\rn_j u_{m-j}
\label{orderhigher}
\end{align}
\label{Lorders}
\end{subequations}
for $m = 2,3,4,\cdots$. The solution $u_0$ is
\begin{equation}
u_0(x) = \disp\sum_{m=-\infty}^{\infty} a_m \re^{\ri(2 m + \rn_0)x},
\end{equation}
where
\begin{equation}
a_m = \frac{(V_0/8)^m}{m! (m+ n_0)!}, \quad \mbox{for}
\hspace{0.2cm} m\ge 0,
\end{equation}
and $a_m=0$ for $m<0$. This solution comes directly from
(\ref{e:Jk}) by replacing the wavenumber $k$ with the integer
$\rn_0$. The remaining linear inhomogeneous equations
(\ref{order1})-(\ref{orderhigher}) for $u_1, u_2, \dots$ will be
solved by first imposing the solvability condition due to the
Fredholm Alternative Theorem and then expanding the solution in
terms of Fourier series.

The adjoint operator of $L$ is
\begin{equation}
L^A = \partial_{xx} + \rn_0^2 +  \frac{V_0}{2} \re^{-2 \ri x},
\end{equation}
and has kernel $u_0^*$ since $L^A u_0^* = 0$. The Fredholm
Alternative Theorem requires that the forcing terms in equations
\eqref{Lorders} be orthogonal to $u_0^*$. As we have mentioned
earlier, we are concerned with $\pi$- and $2\pi$-periodic solutions
here and thus define the inner product as
\begin{equation}
\langle  f(x),g(x)\rangle  = \frac{1}{2\pi}\disp\int_{-\pi}^{\pi}
f(x) g^*(x) \rd x.
\end{equation}
Using the fact that $\langle  \re^{\ri p x}, \re^{\ri q x}\rangle  =
\delta_{p,q}$ for integers $p, q$, we obtain the inner product
\begin{equation}
\langle  u_0 , u_0^*\rangle = \left\{ \begin{array}{ll} a_0^2, &
\quad
\rn_0 = 0, \\
0, & \quad \rn_0 = 1,2,3, \cdots. \end{array}\right.
\end{equation}
Eq. \eqref{order1} for $u_1$ has the solvability condition
\begin{equation}
0 = -\rn_1\langle  u_0 , u_0^*\rangle.
\end{equation}
Thus, when $\rn_0 = 0$ then $\rn_1 = 0$. For other $\rn_0$ this
solvability condition is satisfied automatically and the solution
$u_1$ may be formally written as
\begin{equation}
u_1 = -\rn_{1}L^{-1}u_0.  \label{u1linear}
\end{equation}
Expanding $L^{-1}u_0$ into Fourier series
\begin{equation}
L^{-1}u_0= \disp\sum_{m=-\infty}^{\infty} b_m \re^{\ri(2m+\rn_0)x}
\label{bSum}
\end{equation}
and substituting it into $L \left[ L^{-1} u_0\right] = u_0$ we find
that the coefficients $b_m$ satisfy the recursion relation
\begin{equation}
-4\left(m^2 + m\rn_0\right)b_m + \frac{V_0}{2}b_{m-1} = a_m
\label{LB}
\end{equation}
for all integers $m$. The relevant coefficients are
\begin{subequations}
\begin{align}
b_{-1} & = \frac{2}{V_0} a_0 \\
b_{-2} & = -\frac{16}{V_0^2}(\rn_0-1)a_0\\
\vdots & \nonumber \\
b_{-\rn_0} &= \disp\frac{(-1)^{\rn_0-1}}{4} (\rn_0-1)!^2 \left(\frac{8}{V_0}\right)^{\rn_0} a_0 \\
b_{m} &= 0, \quad \mbox{for} \hspace{0.2cm} m < -\rn_0.
\end{align}
\end{subequations}
Notice that this series also terminates in the negative $m$
direction at $m = -\rn_0$.

The equation \eqref{orderhigher} for $u_2$ is
\begin{equation}  \label{e:Lu2}
L u_2 = -\ri \sin(2x) u_0  - \rn_1 u_1 - \rn_2 u_0.
\end{equation}
When $\rn_0=0$ (hence $\rn_1 = u_1=0$), its solvability condition is
\begin{equation}
\rn_2 = -\frac{\langle  \ri \sin(2x)u_0, u_0^*\rangle  }{\langle
u_0, u_0^*\rangle },
\end{equation}
which gives $\rn_2 = V_0/8$. For $\rn_0 \geq 1$, after substituting
in the solution \eqref{u1linear}-\eqref{bSum} for $u_1$, the
solvability condition of (\ref{e:Lu2}) gives
\begin{equation}
\rn_1^2 = \disp\frac{\langle  \ri \sin(2x) u_0, u_0^*\rangle
}{\langle  L^{-1}u_0, u_0^*\rangle }.
\end{equation}
By rewriting $\ri \sin(2x) = \frac{1}{2}\left(\re^{\ri 2 x} -
\re^{-\ri 2 x}\right)$  we may again use the orthogonality of the
Fourier modes to work out the inner products explicitly,
\begin{subequations}
\begin{align}
\langle  \ri \sin(2x) u_0, u_0^*\rangle  &= -\frac{1}{2}a_0^2, ~~~&{\rm for}& ~ \rn_0 = 1, \\
\langle  \ri \sin(2x) u_0, u_0^*\rangle  &= 0, ~~&{\rm for}& ~ \rn_0 = 2,3,4\ldots,  \label{e:sin2xu0u0} \\
\langle  L^{-1}u_0, u_0^*\rangle  &= b_{-\rn_0} a_0, ~~&{\rm for}& ~
\rn_0 = 1,2,3 \ldots.
\end{align}
\end{subequations}
Thus,
\begin{equation}
\rn_1 = \pm \ri \frac{V_0^{1/2}}{2}, \quad \mbox{for} \hspace{0.2cm}
\rn_0 = 1,
\end{equation}
and $\rn_1=0$ for $\rn_0>1$.  This means if $\rn_0 = 1$ then $\rn_1$
is an imaginary number and, returning to the expansion for $\mu$ in
equation \eqref{muexpand}, that $\mu$ is a complex number for $W_0$
above the phase transition point, $\epsilon>0$, and real below,
$\epsilon<0$. This is the bifurcation that occurs at edge of the
Brillouin zone where the first and second bands merge (see Figs.
\ref{f:fig1} and \ref{f:fig2}).

For $\rn_0=1$, we can proceed to solve Eq. (\ref{e:Lu2}) for $u_2$
by Fourier expansion. Then from the solvability condition for the
$u_3$ equation we can find that
\[ \rn_2=\frac{V_0}{32}, \quad \mbox{for} \hspace{0.2cm}
\rn_0 = 1. \]

For $\rn_0\geq2$ we formally write the solution $u_2$ as
\begin{equation}
u_2 = -L^{-1}\left[ \ri \sin(2x) u_0 \right] - \rn_2 L^{-1}u_0
\label{u2linear}
\end{equation}
since we know that $L^{-1}[\ri \sin(2x) u_0]$ is well defined in
view of the orthogonality (\ref{e:sin2xu0u0}). Expanding it into a
Fourier series
\begin{equation}
L^{-1}\left[ \ri \sin(2x) u_0\right] =
\disp\sum_{m=-\infty}^{\infty} c_m \re^{\ri(2m+\rn_0)x},
\end{equation}
it is easy to find that the coefficients $c_m$ satisfy the recursion
relation
\begin{equation}
-4\left(m^2 + m\rn_0\right)c_m + \frac{V_0}{2}c_{m-1} =
\frac{1}{2}\left(a_{m-1} - a_{m+1} \right),  \label{LC}
\end{equation}
and
\[
c_{-1} = -\disp\frac{1}{8(\rn_0 +1)} a_0,  \quad c_{-2} = -\disp
\frac{2}{(1+\rn_0)V_0}a_0.
\]
Again there are only a finite number of terms in the negative $m$
direction,  i.e. $c_m = 0$ for $m < -\rn_0$.

For $\rn_0 \geq 2$ (hence $\rn_1 =u_1=0$), Eq. \eqref{orderhigher}
for $u_3$ is
\begin{equation}
L u_3 =- \rn_3 u_0,
\end{equation}
thus
\begin{equation}
u_3= -\rn_3 L^{-1} u_0,  \label{u3linear}
\end{equation}
and Eq. \eqref{orderhigher} for $u_4$ is
\begin{equation}
L u_4  =- \ri \sin(2x)  u_2 -\rn_2 u_2 -\rn_4 u_0.
\end{equation}
After substituting in \eqref{u2linear} for $u_2$ the solvability
condition is
\begin{align}
0 =& ~\rn_2^2 \langle  L^{-1}u_0, u_0^*\rangle   \nonumber \\
&+ 2\rn_2 \langle  L^{-1}\ri \sin(2x)u_0, u_0^* \rangle  \nonumber \\
& +   \langle  \ri \sin(2x) L^{-1}\ri \sin(2x) u_0 , u_0^*\rangle,
\label{e:doubleroot}
\end{align}
with coefficients given by
\begin{subequations}
\begin{align}
\langle  L^{-1}u_0, u_0^*\rangle  &= ~b_{-\rn_0} a_0,  \label{e:Linvu0u0} \\
\langle  L^{-1}\ri \sin(2x)u_0, u_0^* \rangle  &= ~c_{-\rn_0} a_0, \\
\langle  \ri \sin(2x) L^{-1}\ri \sin(2x) u_0 , u_0^*\rangle  &=
-\frac{1}{2}\left( c_{-\rn_0}a_1 + c_{-\rn_0+1} a_0 \right).
\end{align}
\end{subequations}
For $\rn_0 = 2$ this gives $\rn_2 = V_0/24$, $- 5V_0/24$. For $\rn_0
\geq 3$ we find that $\rn_2$ is a double root,
\begin{equation}
\rn_2 = - \disp\frac{V_0}{8} \disp\frac{1}{\rn_0^2 -1}.  \label{n2}
\end{equation}
At these $\rn_2$ values, the solution $u_4$ is well defined and is
given by
\begin{equation}
u_4 = - L^{-1}\left[ \ri \sin(2x) u_2 + \rn_2 u_2 \right] - \rn_4
L^{-1} u_0.
\end{equation}

For $\rn_0 \geq2$, Eq. \eqref{orderhigher} for $u_5$ is
\begin{equation}
L u_5 =- \ri \sin(2x) u_3 - \rn_3 u_2 - \rn_2 u_3 -\rn_5 u_0.
\end{equation}
After substituting in equations \eqref{u2linear} and
\eqref{u3linear} the solvability condition for this $u_5$ equation
reduces down to
\begin{equation} \label{e:n2n3}
0 = \rn_3 \left[ \rn_2 \langle  L^{-1}u_0, u_0^*\rangle  + \langle
L^{-1}\ri \sin(2x)u_0, u_0^* \rangle \right].
\end{equation}
Thus, for $\rn_0 = 2$ we must have $\rn_3 = 0$; and for $\rn_0 \geq
3$ this condition is satisfied automatically since $\rn_2$ from Eq.
\eqref{n2} is a double root of Eq. (\ref{e:doubleroot}).

For $\rn_0 \geq3$, Eq. \eqref{orderhigher} for $u_6$ is
\begin{equation}
Lu_6  = - \ri \sin(2x) u_4 -\rn_2 u_4 -\rn_3u_3 -\rn_4 u_2 - \rn_6
u_0.
\end{equation}
Substituting in Eq. \eqref{u3linear} and noting that $\langle
u_2,u_0^*\rangle =0$ (in view of (\ref{e:n2n3})) we are left with
the solvability condition
\begin{equation}  \label{e:n3}
\rn_3^2 = \disp\frac{\langle  u_4 (\rn_2 + \ri \sin(2x)) ,
u_0^*\rangle }{\langle  L^{-1} u_0, u_0^*\rangle }.
\end{equation}
This condition may be further simplified,
\begin{align*}
\langle  u_4 (\rn_2 &+ \ri \sin(2x) ), u_0^*\rangle    \\
&= \langle  u_4 , (\rn_2 - \ri \sin(2x)) u_0^*\rangle   \\
& = - \langle   \ri \sin(2x) u_2 + \rn_2 u_2 - \rn_4 u_0, u_2^*\rangle  \\
& = - \langle  \ri\sin(2x) u_2 , u_2^*\rangle.
\end{align*}
Thus for $\rn_0=3$, we get
\begin{equation}  \label{e:n3n03}
\rn_3 = \pm\ri \disp\frac{ ~V_0^{3/2}}{2^9},
\end{equation}
and for $\rn_0>3$, we get $\rn_3=0$. This shows that there is
another bifurcation point of complex eigenvalues at $\rn_0=3$, where
the third bandgap closes (see Figs. \ref{f:fig1} and \ref{f:fig2}).

The results of the above perturbation calculations are summarized in
Table 1 of the main text. Continuing these calculations to higher
$\rn_0$ values, we have found that the coefficient $\rn_{2m+1}$ is
always imaginary for $\rn_0=2m+1$, where $m=0, 1, 2, \cdots$. Thus
complex eigenvalues bifurcate out simultaneously from $\rn_0=1, 3,
5, \cdots$ at the phase transition point $W_0=0.5$.

\section*{Appendix 2: Analytical criterion for zero-eigenvalue bifurcation of solitons in complex potentials}

In real potentials (such as when $W_0=0$ in (\ref{e:PTlattice})),
the power curve does more than just a convenient way to catalogue
and parameterize a continuous family of solitons for various values
of the propagation constant $\mu$. Specifically, whenever the power
curve has a local extremum the zero eigenvalue in the
linear-stability spectrum of solitons then bifurcates out along the
real and imaginary axes on the two sides of the power extremum
respectively, leading to a change of stability at the power extremum
(if no other unstable eigenvalues exist) \cite{Yang_book}. In this
appendix we consider the extension of this concept to general
complex potentials (which include PT-symmetric lattices as special
cases). The resulting analytical criterion for zero-eigenvalue
bifurcation will explain the stability switchings in Fig.
\ref{PowerCurve1DDefocusing} and Fig. \ref{PowerCurve2D} (right
side), as well as the onset of real eigenvalues in Fig.
\ref{Spectrum1DFocusing} (right panel).

Let us begin with the eigenvalue problem \eqref{EigenProblem}
derived in the main text,
\begin{equation}
\ri {\cal L} \left(\begin{array}{c} \tilde{u}\\ \tilde{w}
\end{array} \right) = \lambda   \left(\begin{array}{c} \tilde{u}\\ \tilde{w}
\end{array}\right),  \label{EigProblemSS}
\end{equation}
where we know that $\lambda= 0$ is always an eigenvalue with
algebraic multiplicity of at least two due to phase invariance of
Eq. (\ref{e:model}). The eigenfunction and generalized eigenfunction
of this zero eigenvalue associated with the phase invariance can be
written explicitly in terms of the soliton $u(\textbf{x})$,
\begin{align*}
{\cal L} \vectorTwo{u}{-u^*}  =0 ~~~~~ {\rm and}~~~~~ {\cal L}
\vectorTwo{u_{\mu}}{u_{\mu}^*} = \vectorTwo{u}{-u^*}.
\end{align*}
Thus, for nonzero eigenvalues to bifurcate out from the origin,
$\lambda=0$ must have algebraic multiplicity of at least 3 at that
point. A sufficient condition for this to occur is that there be a
second generalized eigenfunction $\psi$ which solves
\begin{equation}
{\cal L} \psi = \vectorTwo{u_{\mu}}{u_{\mu}^*}.  \label{Eigen3}
\end{equation}
We now use the Fredholm Alternative Theorem to derive the
solvability condition for Eq. \eqref{Eigen3}. Let us denote the
kernel of the adjoint operator ${\cal L}^A$ as $\phi^{(A)}$, i.e.,
\begin{equation}
{\cal L}^A \phi^{(A)} = 0,
\end{equation}
where the adjoint operator is
\begin{equation}
{\cal L}^A = {\cal L}^{*T}.
\end{equation}
Here the superscript ``$T$" stands for transpose of a matrix.
Then the solvability condition of Eq. (\ref{Eigen3}) is
\begin{equation}
\left\langle \vectorTwo{u_{\mu}}{u_{\mu}^*}, \,   \phi^{(A)}
\right\rangle =0,  \label{e:criterion}
\end{equation}
which is a sufficient condition (criterion) for zero-eigenvalue
bifurcation in general complex potentials.

For real potentials, it is easy to see that
\begin{equation}
\phi^{(A)} = \left( \begin{array}{c} u(x) \\  u^*(x)
\end{array} \right),
\end{equation}
thus the above criterion reduces to $P'(\mu) = 0$, i.e., the
extremum of the power curve \cite{Yang_book}. For general complex
potentials, however, $\phi^{(A)}$ is not equal to the above
expression, thus stability switching will no longer occur at a power
extremum. An example of this has been seen in Fig.
\ref{PowerCurve1DDefocusing}.

\end{document}